\documentclass[12pt]{book}
\usepackage{plenum,epsfig}
\begin{document}

\chapter{TOPOLOGICAL CONTENTS OF 3D SEIBERG-WITTEN THEORY}

\author{Bogus\l aw Broda}

\affiliation{Department of Theoretical Physics\\
University of \L\'od\'z\\
Pomorska 149/153\\
PL--90-236 \L\'od\'z\\
Poland}

\section{INTRODUCTION}

In three dimensions (3D), we have the two interesting topological quantum field theories of ``cohomological'' type:  $SU(2)$ non-abelian topological gauge theory of flat connection and 3D version of the (topological) Seiberg-Witten (SW) theory. The former is a 3D twisted  $SU(2)$ pure gauge ${\cal N}=4$ SUSY theory or a 3D version of the Donaldson-Witten (DW) theory, and by ``definition'' it describes the Casson invariant, which appropriately counts the number of flat $SU(2)$ connections.\refnote{\cite{bt}} The latter is a 3D twisted version of abelian ${\cal N}=4$  SUSY theory with a matter hypermultiplet.\refnote{\cite{w2}} Also that theory should describe an interesting non-trivial topological invariant of 3D manifolds pertaining to SW invariant, which was conjectured to be equivalent to the Casson invariant (and thus to the former theory), and also to topological torsion.\refnote{\cite{mthl}} The first conjecture is physically strongly motivated by the fact that the both theories can be derived from 4D  $SU(2)$ pure gauge ${\cal N}=2$ SUSY theory corresponding via twist to DW theory.\refnote{\cite{w1}} That equivalence would be a 3D counterpart of the equivalence of 4D DW and SW theories. The latter being a ``low-energy version'' of the former. In this paper we present a qualitative physical scenario supporting the conjecture that topological contents of 3D SW theory is basically equivalent to an abelian version of the Casson invariant (of an auxiliary space). In turn, the abelian Casson invariant will be shown to be equivalent to the Alexander ``polynomial'' of the manifold.\refnote{\cite{fn}} In fact, as follows from mathematical literature the Alexander invariant is connected to the (non-abelian) Casson invariant as well as to topological torsion.\refnote{\cite{t}}

\section{PHYSICAL SCENARIO}

Our physical inspiration is coming from an intuitive idea of Taubes and Witten\refnote{\cite{w3}} concerning ``superconducting phenomena'' in 4D SW theory. In mathematics, its classical version gives rise to the, so-called, ``vanishing theorems''.\refnote{\cite{w2}} Using physical arguments (Higgs mechanism, superconductivity, infrared regime, duality) and a geometric-topological construction (scalar curvature distribution compatible with surgery), we propose a topological interpretation of 3D SW theory in terms of the abelian Casson invariant. Further algebraic reasoning shows equivalence of that invariant to the Alexander ``polynomial''. The scenario involves several simple steps. Our starting point is a 3D version (a dimensional compactification) of the original SW theory. Observing that the scalar curvature $R$ plays the role of a mass-squared parameter for the monopole field (in the original 4D case as well as in our 3D one) we can use that observation to control the theory in low-energy limit. In the generic case of an arbitrary closed connected 3D manifold ${\cal M}^3$, there are regions ${\cal M}_+^3$ of positive scalar curvature, $R>0$, and regions ${\cal M}_-^3$  of negative scalar curvature, $R<0$ (we discard the regions of zero scalar curvature, $R=0$). In fact, we will be able to assume that ``almost'' the whole manifold ${\cal M}^3$ is of positive scalar curvature. Roughly speaking, we have a 3D manifold of positive scalar curvature with ``bubbles'' of negative curvature in the form of (in general, knotted) ``superconducting'' circuits/rings. In  ${\cal M}_+^3$, the matter is massive, and (super)electromagnetostatics is a long-range interaction, whereas in ${\cal M}_-^3$ we deal with the Higgs mechanism and the fields become short-range. In other words, due to the Meissner effect electromagnetic fields are expelled from the superconducting regions ${\cal M}_-^3$ (see, Fig.~1).
%	Fig.1
\begin{figure}[t]
\centering
\epsfig{file=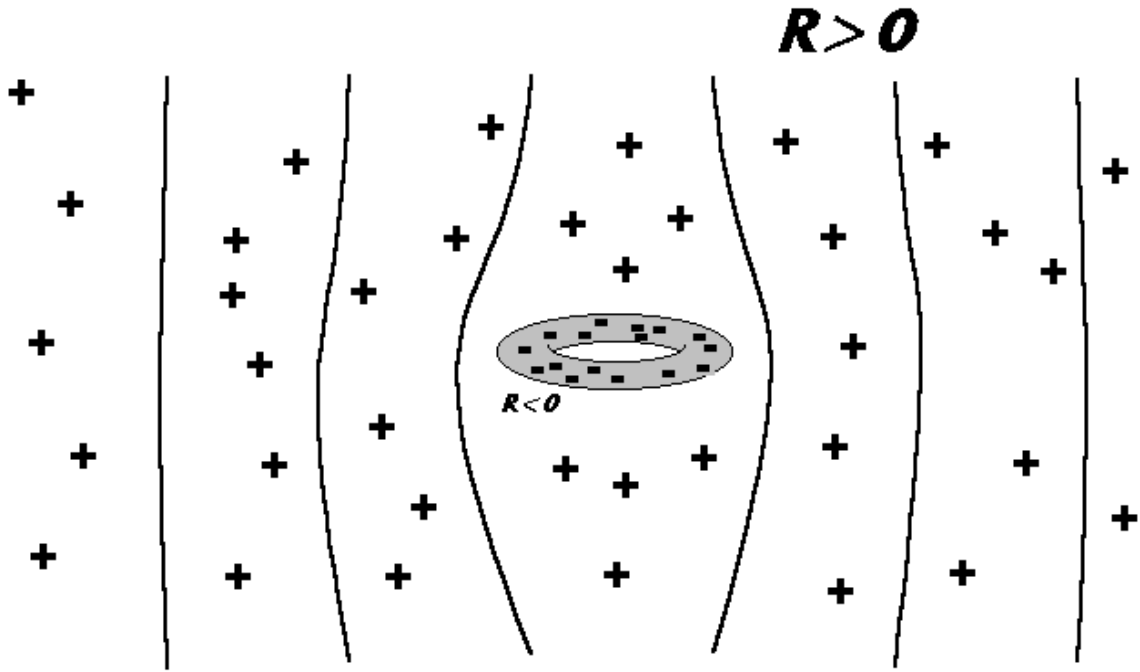, width=15cm}
\caption{All fields become expelled from the (superconducting) ring of negative scalar curvature.}
\end{figure}
In low-energy limit, we obtain pure (super)electromagnetostatic fields living only in ${\cal M}_+^3$. In topological sector, due to abelianity of the theory, the fields could detect the first cohomology of ${\cal M}_+^3$ (rather a trivial quantity). But happily, that is not the case as the theory is ill-defined in original variables, or at least not reliable because the coupling constant diverges. The trick of Seiberg and Witten to dualize the variables improves the situation. However, there is some subtlety in the case of non-simply-connected manifolds. We know from classical electrodynamics that, for example, the magnetostatic potential lives on the infinite cyclic covering space $\widetilde{{\cal M}^3}$ of the original manifold ${\cal M}^3$. Therefore, we begin to measure the first (co)homology of $\widetilde{{\cal M}_+^3}$ (the Alexander invariant), rather than trivial one of ${\cal M}_+^3$. Strictly speaking, we count the number of flat abelian connections (on $\widetilde{{\cal M}^3}$), an abelian version of the Casson invariant, which we next ``explicitly calculate'' obtaining the Alexander ``polynomial''.
What is left to show is a purely geometric-topological relation between the distribution of the scalar curvature $R$ of ${\cal M}^3$ and a topological description of ${\cal M}^3$. It appears that it is possible to establish a necessary correspondence via surgery.

It is very instructive and suggestive to graphically present a web of interesting mutually related topological field theories in various dimensions (3 and 4 in our case) (see, Fig.~2, for a ``three-dimensional'' illustration):
%	Fig.2
\begin{figure}[t]
\centering
\epsfig{file=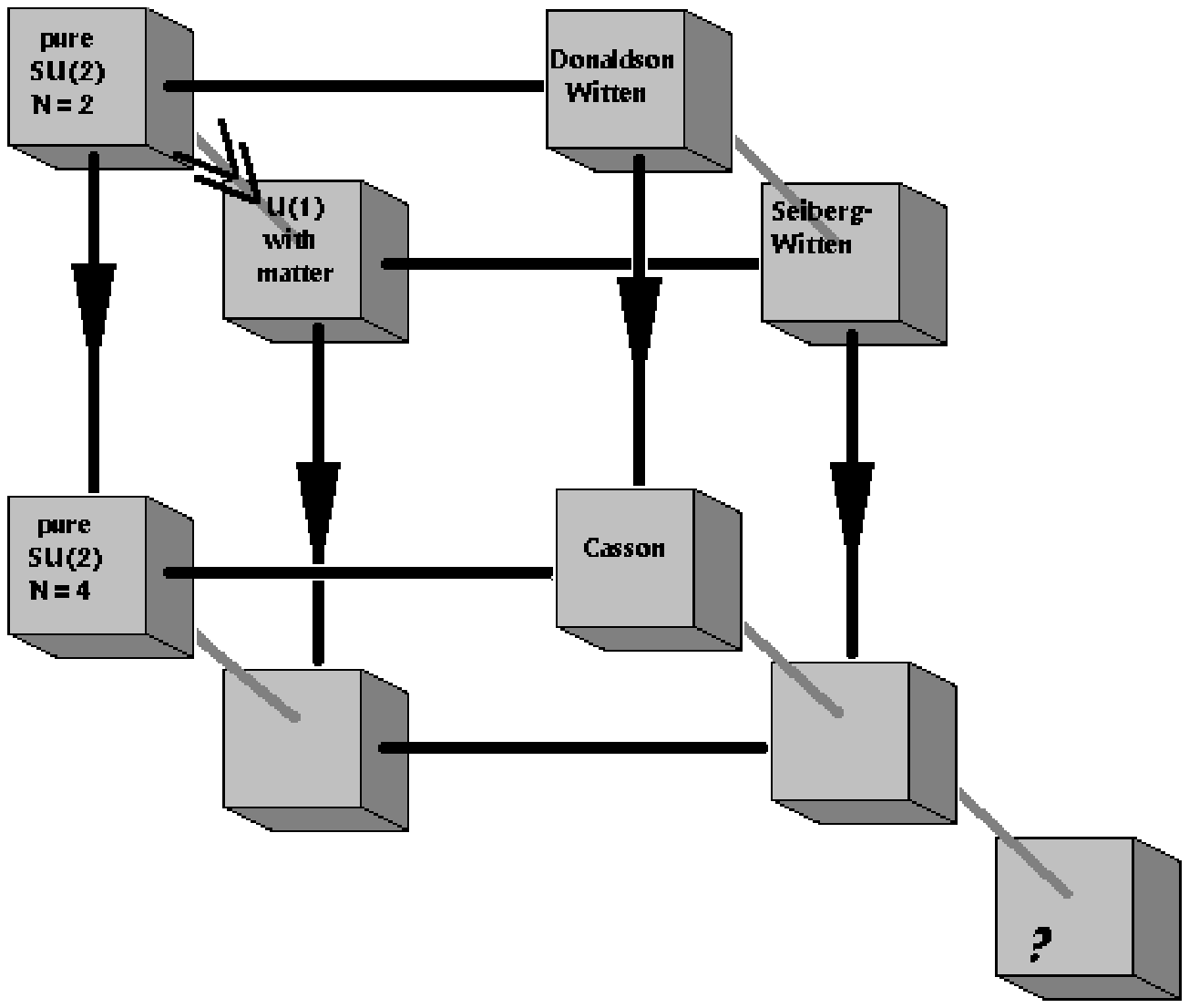, width=15cm}
\caption{Web of mutually related quantum field theories in 3D and 4D.}
\end{figure}
on the left-hand side we have ``physical'' theories, and on the right-hand side, the corresponding twisted (topological) ones; at the top, the dimension $D=4$, whereas at the bottom, $D=3$; in front, we have primary, microscopic theories, behind their low-energy limits (effective ones); solid black lines denote twists; solid black lines with arrows---dimensional reduction/compactification; a grey line with arrows---renormalization group procedure; other grey lines represent implicit or suspected connections between theories. The theory of interest in this paper could stand somewhere in the place of the box with the question mark. Since dimensional compactification/reduction and renormalization group procedures are not commutative, a further low-energy limit we perform should make sense (therefore, some objects at the bottom of Fig.~2 should be a bit fuzzy).

\section{SUPERCONDUCTIVITY AND DUALITY}

There are two equivalent possibilities to reach 3D version of SW theory. Either we can dimensionally compactify 4D SW theory or we can twist 3D (dimensionally compactified) abelian ${\cal N}=4$  SUSY theory coupled to monopole field. The ``classical'' action of 4D SW theory, which we could assume as our starting point, looks as follows\refnote{\cite{w2}}
$$
S=\int_X d^4x \sqrt{g} \left( {1\over2}|F^+|^2 + g^{\mu\nu} D_\mu M^A D_\nu \overline{M}_A
+ {1\over2}|M|^4 + {1\over4}R|M|^2 \right),
\eqno{(1)}
$$
where $F$($=dA$) is the strength of electromagnetic field, and $M$ is a monopole field. The monopole field $M$ is initially massless but in curved space the response of that field to gravitational interactions yields the scalar curvature $R$ which effectively plays the role of the mass-squared parameter. Upon dimensional compactification on $X={\cal S}^1\times Y$, where ${\cal S}^1$ is a circle (of radius 1, for simplicity), we have the following decomposition of the $U(1)$ gauge field
$$
A_\mu \longrightarrow (\phi, A_i).
$$
Besides ordinary gauge transformations of $A_i$ field, there is a residual gauge transformation of $\phi$ field,
$$
\phi\rightarrow \phi+1,
\eqno{(2)}
$$
which follows from the formula
$$
A_0 \longrightarrow A_0 - i\partial_0 e^{ix_0}.
$$
Thus, $\phi$ field assumes values in ${\cal S}^1$ rather than in $R^1$.

As we have already mentioned, in generic case, the 3D manifold ${\cal M}^3$ consists of the three kinds of regions: (1) ${\cal M}_+^3$ with $R>0$; (2) ${\cal M}_-^3$ with $R<0$, and (3) ${\cal M}_0^3$ with $R=0$. The third kind is marginal (lower-dimensional), and we will ignore it. In ${\cal M}_+^3$, the theory is in the Coulomb phase---effective theory contains only (free) pure (electro)magnetic long-range interactions---the matter fields are effectively massive and decouple from the theory in the infrared regime. In ${\cal M}_-^3$, the theory is in the Higgs phase, and there are no long-range interactions at all. In other words (SUSY) (electro)magnetic interactions are repelled from ${\cal M}_-^3$. In solid-state terminology, we have superconducting probes ${\cal M}_-^3$ put in ${\cal M}^3$. What makes the analogy even closer and more suggestive is the fact that the dimension ($=3$) is physical and the shape of the probes is ``realistic'': they are (possibly, knotted) circuits/rings, as will be explained later on. The difference lies in supersymmetric extension and in our limitation to topological sector. Thus, in the infrared limit we have only pure SUSY electromagnetic interactions which penetrate ${\cal M}_+^3$. In twisted, topological sector it could naively correspond to measuring the first cohomology $H^1({\cal M}_+^3)$, not a very exciting invariant.

In 3D the coupling constant $e$ is of positive dimension (${\rm dim}\,e=m^{1/2}$), and therefore it diverges in the infrared limit. But there is a well-known procedure to rescue the theory, namely we can pass to dual variables inverting the coupling constant $e$, making the theory more convergent. The dualization we propose is not the standard 3D one, but it is ``four-dimensional''.\refnote{\cite{sw}} It means that one of the components of the scalar part of the supermultiplet is dualized tensorially, i.e.\ to a vector, as it were an electrostatic potential of a 4D theory. It is sensible because we can treat our 3D theory as a time-independent 4D one since we think in terms of compactification on $X$($={\cal S}^1\times Y$). Such a dualization does ``almost nothing'' to the variables, the scalar and vector potentials become simply exchanged
$$
A_i \longrightarrow \chi,
$$
$$
\phi \longrightarrow V_i.
$$
``Almost nothing'' means that although the dual potentials have the same tensorial structure as the original ones, they are multivalued.
On a simply-connected manifold nothing really topologically interesting happens. In general case however, since the dual potentials are multivalued functions, they begin to live on an infinite cyclic covering space of a given 3D manifold. That way, in topological sector, the theory switched from measuring the first cohomology of ${\cal M}_+^3$ to measuring the first cohomology of its infinite cyclic cover, $\widetilde{{\cal M}_+^3}$. In other words, upon twisting, we have an abelian ``cohomological'' theory on $\widetilde{{\cal M}_+^3}$.
It should be stressed that the conclusion that our theory lives on a covering space as a result of the multivaluedness of fields can be true only in topological sector as only this sector uses ``constant'' configurations for which the conclusion is satisfied. Since ${\cal M}_+^3$ has a boundary it would be rather difficult, if not impossible, to explicitly construct a local action for both the supersymmetric and even twisted version of the theory.
But it is not necessary, since we are able to implicitly identify the theory (the fields are still known explicitly) in topological sector. Namely, it is a theory of abelian flat connection on $\widetilde{{\cal M}_+^3}$. The partition function $Z$ of our theory should ``ordinarily'' (without sign, due to abelianity) count the number of (inequivalent) $U(1)$-representations of the fundamental group $\pi_1$ of $\widetilde{{\cal M}_+^3}$
$$
Z = \# \hbox{Rep} \left( \pi_1\left(\widetilde{{\cal M}_+^3}\right) \longrightarrow U(1)\right).
\eqno{(3)}
$$
Eq.~(3) defines an abelian analog of the Casson invariant. Our temporary conclusion says that the partition function of 3D SW theory describes the abelian Casson invariant $Z$ of $\widetilde{{\cal M}_+^3}$.

\section{SHAPE OF THE SUPERCONDUCTING PROBES}

We have already suggested that the superconducting probes ${\cal M}_-^3$ are of the form of (in general, knotted)  circuits/rings. More precisely, they are thicken knots (links). Mathematically, their relation to ${\cal M}^3$ is the most natural, standard one---via the, so-called, surgery procedure---${\cal M}^3$ should be considered as a manifold obtained according to surgery instructions encoded in the knot/link, i.e.\ via cutting out (thicken) knots and pasting them back in a different way.

For our purposes we need a more or less obvious statement which, in a sense, is  complementary to the Gromov-Lawson-Schoen-Yau (GLSY) theorem.\refnote{\cite{glsy}} The GLSY theorem concerns, roughly speaking, higher-dimensional situation, and it says that carefully performing surgeries on a manifold of positive scalar curvature we do not spoil the positivity property. In our (lower-dimensional) case the GLSY theorem happily fails in generic case, and we expect something opposite, i.e.\ knots can introduce regions of negative scalar curvature. To see it (see, Fig.~3), let us consider a (straight) cylinder instead of a (curved) thicken knot for a moment. Cutting it vertically with a plane, i.e. discarding the third dimension, we obtain a circle on the plane. Adding the ``fourth'' dimension as a third one we can attach a ``pipe'' (a dimensionally reduced handle) to the plane along the circle. After smoothing out, the attaching area has obviously negative curvature.
%	Fig.3
\begin{figure}[t]
\centering
\epsfig{file=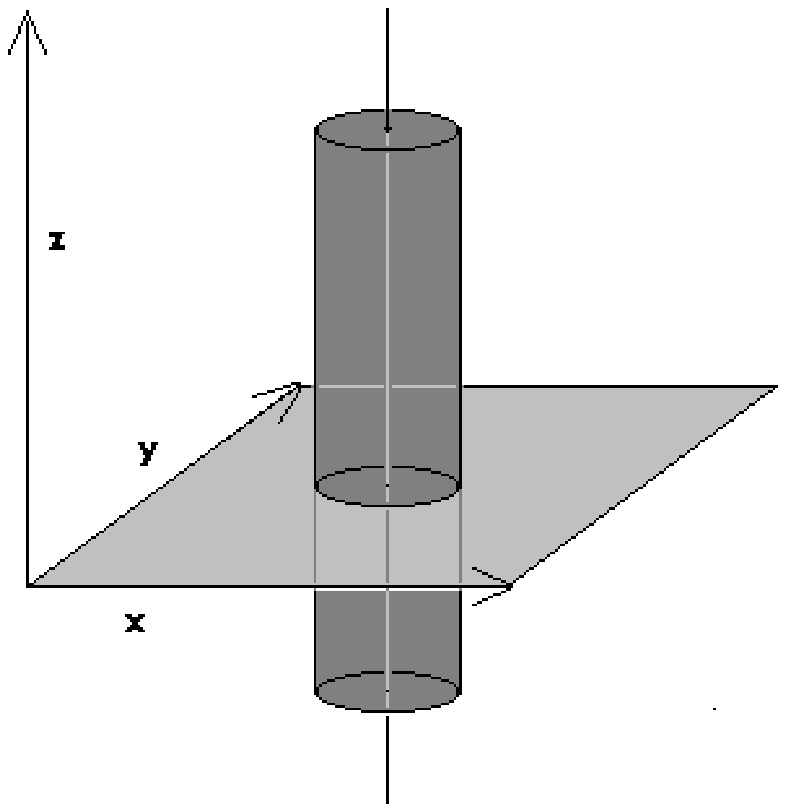, width=7cm, clip=}
\epsfig{file=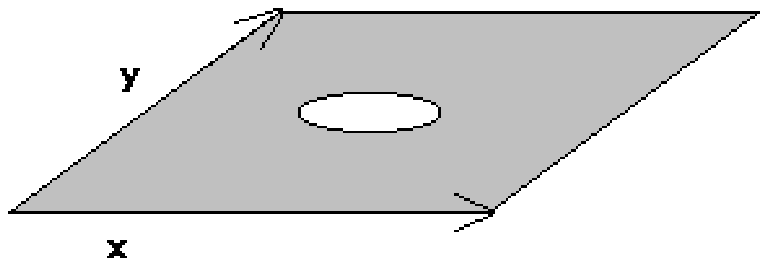, width=7cm, clip=}
\epsfig{file=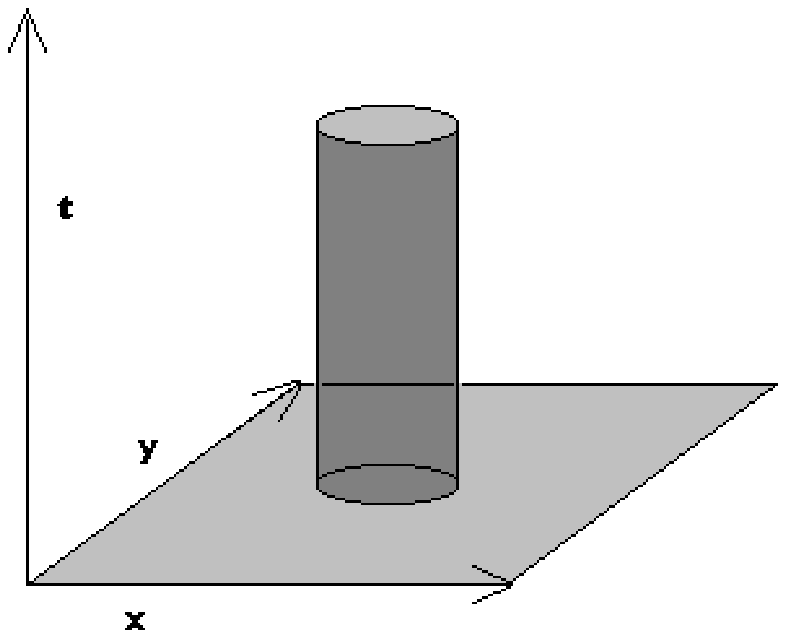, width=7cm, clip=}
\epsfig{file=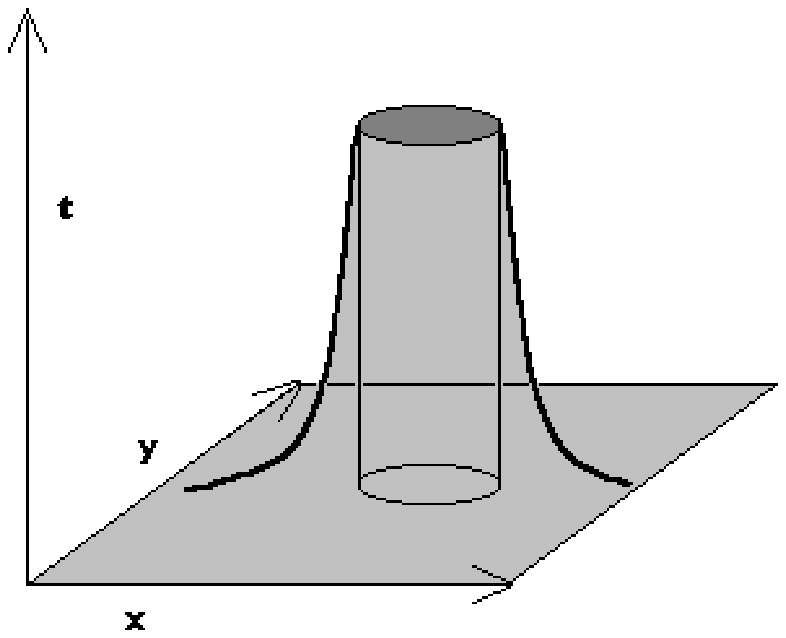, width=7cm, clip=}
\caption{Scalar curvature compatible with surgery.}
\end{figure}
Cartesian product with the ``third'' (temporarily discarded) dimension does not change the sign of the scalar curvature. In reality, a knot is curved rather than straight but since we are interested in topology we can arrange the curvature of the attaching area large enough in comparison with the curvature of the knot itself.

\section{CALCULATING THE PARTITION FUNCTION}

Finally, we would like to know what is the relation between $Z$ and some, possibly known topological invariant of ${\cal M}^3$. To this end, we will ``explicitly calculate'' the number of $\hbox{Rep} \left( \pi_1\left(\widetilde{{\cal M}_+^3}\right) \longrightarrow U(1)\right)$. First of all, since $U(1)$ is an abelian group a non-abelian part of $\pi_1$ drops out, and effectively we deal with an equivalent but simpler expression,
$$
Z=\# \hbox{Rep} \left( H_1\left(\widetilde{{\cal M}_+^3}\right) \longrightarrow U(1)\right),
\eqno{(4)}
$$
where $H_1$ is an integer-valued first homology group. We claim that $Z$ is the Alexander ``polynomial'' $A$ understood as
$$
A=\det A_{kl},
\eqno{(5)}
$$
where
$$
\sum_l A_{kl} \alpha_l=0,
\qquad
k,l=1,\dots,N,
\eqno{(6)}
$$
for $\alpha_l$---some homology basis. Eq.~(6) represents the homology we are interested in, i.e.\ $H_1\left(\widetilde{{\cal M}_+^3}\right)$ (a mathematical fact). Now, we should pass to a $U(1)$ representation of (6). If
$$
\alpha_l \longrightarrow e^{2\pi i \omega_l},
$$
then for LHS of (6) we have
$$
\sum_l A_{kl} \alpha_l \longrightarrow \prod_l \exp\left(2\pi i A_{kl} \omega_l\right)
= \exp\left(2\pi i \sum_l A_{kl} \omega_l\right),
$$
where $0\leq\omega_1,\dots,\omega_N<1$, for uniqueness. RHS of Eq.~(6) is now
$$
0\longrightarrow 1 = \exp(2\pi i m_k),
\qquad
m_k \in Z.
$$
A new, $U(1)$ version of Eq.~(6) is then
$$
\sum_l A_{kl} \omega_l = m_k.
\eqno{(7)}
$$
$Z$ is the number of solutions of Eq.~(7), i.e.\ the number of different $m_k$. $m_k$ are integer-valued points in a parallelepiped spanned by the ``base vectors'' $A_{k(l)}$. One can observe that the number of $m$'s is equal to the volume of that parallelepiped (see, Fig.~4), 
%	Fig.4
\begin{figure}[t]
\centering
\epsfig{file=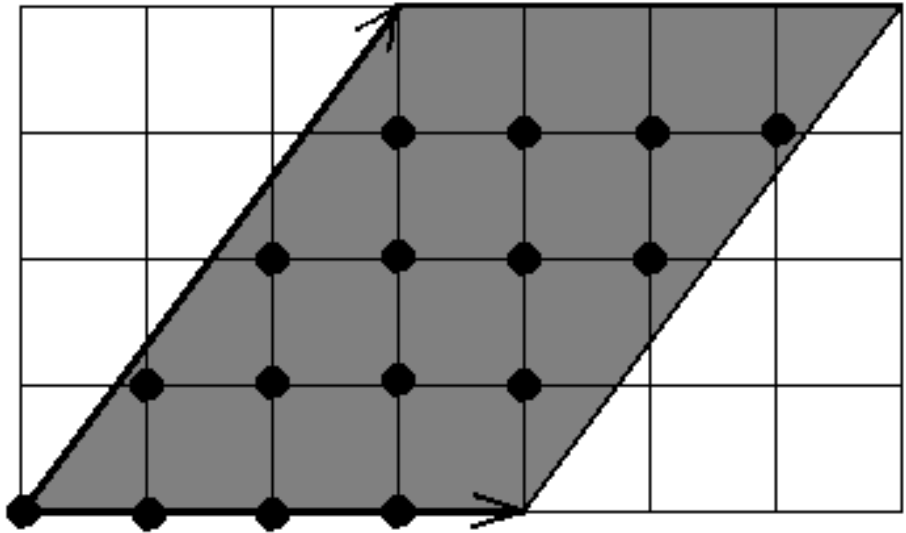, width=12cm}
\caption{Partition function as the number of dots.}
\end{figure}
which in turn, is equal to the determinant $A$ of $A_{kl}$.

\section{CONCLUSIONS}

We have found a physical scenario relating 3D SW theory to an abelian version of the Casson invariant $Z$ of the infinite cyclic covering space $\widetilde{{\cal M}_+^3}$ of ${\cal M}_+^3$. In physical language, the space ${\cal M}_+^3$ emerges as a complement of superconducting probes. Whereas in mathematical language, ${\cal M}_+^3$ emerges as a complement of a thicken knot used to construct the manifold ${\cal M}^3$ via surgery. In low-energy limit, the both languages match due to the Higgs mechanism generated by the scalar curvature $R$. Moreover, the partition function $Z$ has been shown to be equivalent to the Alexander ``polynomial'' of ${\cal M}^3$, which in principle, is related to topological torsion.\refnote{\cite{t}}

\section{ACKNOWLEDGMENTS}

I am greatly indebted to the organizers of the Workshop, Prof.~Poul H.~Damgaard and Prof.~Jerzy Jurkiewicz, for their kind invitation to Zakopane.
The paper has been supported by the KBN grant 2P03B09410.

\end{document}